\newcommand{\un}{\underline}
\newcommand{\fo}{\footnote}
\newcommand{\no}{\normalsize}
\newcommand{\beq}{\begin{quote}}
\newcommand{\enq}{\end{quote}}
\documentclass[11pt]{amsart}
\usepackage{geometry}                
\usepackage{graphicx}
\usepackage{amssymb}
\usepackage{epstopdf}
\usepackage{lineno}
\title{The Reception of Newton's  {\it Principia} }
\author{Michael Nauenberg\\University of California, Santa Cruz}
\begin{document}
\begin{abstract}
Newton's  {\it Principia}, when it appeared in 1687,  was received with the  greatest admiration,  not only by the foremost
mathematicians and astronomers in Europe,  but also by philosophers  like Voltaire and Locke and by  members of
the educated public. In this account I describe some of  the controversies that it provoked, and the impact it had during  the  next 
 century  on the development of  celestial mechanics,  and the theory of gravitation.
\end{abstract}

\maketitle

\section* {Introduction}


 The first edition of Newton' s  
 {\it Mathematical Principles of Natural Philosophy},
commonly known as the {\it Principia}, appeared in 1687,  transforming 
our  understanding of celestial mechanics and gravitational theory. In his magisterial book, 
Newton gave  a physical description  and mathematical proof
for the origin of Kepler's three laws for planetary motion. These laws were empirical laws
 based on the careful observations of the Danish astronomer 
Tycho Brahe, which summarized  most of the  astronomical knowledge about planetary motion that
was known  at the time. 
To accomplish this  feat, Newton  assumed the validity of the principle of inertia 
which he formulated in the { \it Principia}  as the first law of motion \fo{ \no The principle of inertia was formulated
                                           in the form:
                                            \beq
                                             Every body perseveres in
                                             its state of being at rest  or of moving uniformly  straight forward 
                                             except insofar as it is compelled
                                           to change its state by forces impressed (Cohen 1997, 416).
                                           \enq
                                           }, and he
introduced  two fundamental concepts:   that the main attractive gravitational
force  acting on  a planet is a central force  directed towards the 
sun, and  that the  magnitude of this force  is  proportional to the planet's acceleration\fo{\no These  concepts had been formulated
also  by Robert Hooke  who
discussed his ideas with Newton in a lengthy correspondence in 1679  (Turnbull 1960. 297-314) (Nauenberg 1994) (Nauenberg 2005).}. 
From Kepler's third  law\fo {\no The square of the period of the planet  
is proportional to the cube of its  distance from the sun.}, and the approximation that 
planets move with  uniform  motion in a circular orbit,   Newton then deduced that 
this gravitational force
varied inversely with the square of the  radial distance of a planet from the sun.
In addition, Newton proposed that the magnitude of the gravitational force between any two 
bodies was proportional to the product of their {\it inertial} masses, and  by applying Kepler's third law to
the motion  of the moon and the satellites of Jupiter and 
Saturn, he 
of determined  the  masses of these planets and the mass of the earth relative to the mass of the Sun 
\fo{\no This result  was  found among  some early papers by Newton  (Herivel 1965, 196).
                                 Newton and, independently, Christiaan  Huygens, had deduced that  a body moving 
                                with uniform velocity $v$  in a
                                circular orbit  of radius $r$,  has a  radial acceleration $a$ towards the center of the orbit given by $a=v^2/r$.
                                Since   $v=2\pi r/T$, where $T$ is the period,  he obtained  $a=4\pi^2 r/T^2$. Substituting for $T$  Kepler's third law
                                for planetary motion,
                                $T^2=c r^3$, where $c$ is a constant, gives $a=(4\pi^2/c) (1/r^2)$.  In this way Newton found that the acceleration $a$  
                                depends inversely on the square of the radial distance $r$ between a planet and the sun.
                                
                                In the {\it Principia}, Newton proposed that the  same relations apply also to the motion of the 
                                satellites around a planet. 
                                According to his  principle of universal gravitation, $a$ is 
                                proportional to the mass $M$ of the center of force, and therefore
                                $a=GM/r^2$, where $G$ is a universal constant,  now called Newton's constant. 
                                Hence $M=4\pi^2/Gc$, where $c$ is Kepler's constant, and by determining the                              
                                value of $c$ for the satellites of Jupiter, Saturn and the earth,
                                Newton  obtained the ratio  of the mass of each of these planets  
                                relative to the mass of the sun,  given in Prop. 8, Book 3 of the {\it Principia}.
                                }.  
In his essay {\it La cause de la pesanteur}, the great  Dutch scientist Christiaan Huygens remarked
that he was very pleased to read how Newton, by
\beq 
supposing the distance from the earth to the sun to be known, had been able to compute
the gravity that the inhabitants of Jupiter and Saturn would feel compared with what we feel 
here on earth, and what its measure would be on the Sun (Cohen 1997, 219).
\enq

The importance of these  developments was appreciated not only by astronomers and
mathematicians who read the {\it Principia},  but also by   philosophers and by the educated public. The
French philosopher, Fran\c{c}ois Marie Voltaire encapsulated this  recognition with  a succinct comment,
\beq
Avant Kepler tous les hommes etoient aveugles, Kepler fut
borgne, et Newton a eu deux yeux 
\footnote {Before Kepler all men were blind; Kepler was one-eyed, and Newton had
two eyes (Feingold  2004,99)}(Besterman 1968, 83)
\enq
and shortly after Newton's death the English poet Alexander Pope wrote
\beq
Nature, and Nature's Laws lay hid by night\\
God said, let Newton be! and all was light.
\enq

As the reputation of the {\it  Principia} grew, even people who had little or no mathematical ability
attempted to understand its content.  The English philosopher John Locke, who was in exile in Holland,
 went to see Huygens
who assured him of the validity of Newton's propositions.  He was able to follow its conclusions, and later 
befriended Newton, referring  to him ``as the incomparable Mr.Newton"  in the preface of  his essay 
{\it  Concerning Human Understanding} (Locke 2001,13).  While  in exile in England, Voltaire  became acquainted  with Newton's work,
 and after his return to France
he wrote  the {\it  Elemens de la Philosophie de Neuton}  which  popularized Newton's ideas in France.
In this enterprise he was fortunate to have the collaboration of a  gifted companion, Gabrielle Le Tonnelier de
Breteuil, better known as the Marquise du Ch\^{a}telet, who  translated the { \it Principia} into French.
Francesco Algarotti, who was in communication  with Voltaire,   published  his  {\it Newtonianismo per le dame}  which became fashionable in Italy (Feingold 2004).

Initially,  there was  considerable reluctance to accept
Newton's general principles, particularly because  an action at a distance
was  generally  regarded  as due to occult forces,  in  contrast  to  contact 
forces.  According to Descartes,  gravitational forces were due to vortices of unobserved
celestial dust, and  this  explanation had been accepted  by  most Continental
astronomers. At the end of  Book 2 of the {\it Principia}, Newton gave a proof that Cartesian 
vortices were  incompatible with Kepler's second and third laws for planetary motion, but his proof
was based on somewhat arbitrary assumptions about the frictional properties of these vortices, and 
 in an essay, `Nouvelles pens\'{e}e sur le syst\`{e}me de M. Descartes',
the Swiss mathematician Johann Bernoulli gave several objections to this proof (Aiton 1995, 17).
In his {\it Discourse sur
les differentes  figures des astres},  Pierre-Louis Moreau de  Maupertuis  openly defended  Newton's views,  
pointing out  its predictive power,  and remarked that Cartesian impulsion 
was no more intelligible than Newtonian attraction (Aiton 1995, 19), but that universal gravitation
 was  `` metaphysically viable and mathematically superior''
to vortices as an explanation of celestial mechanics (Feingold 2004, 98).  In a remarkable tour de force,
Newton had applied his gravitational theory to determine the shape of the earth,  and he
found that the centrifugal force due to the daily rotation about its axis deformed the earth 
into an oblate spheroid flattened at the poles\fo{\no Assuming that the earth can be regarded as a rotating fluid  
                                 held together by its own attractive gravitational
                                  forces, in  Prop. 19, Book 3, Newton gave a proof that the shape of the earth is an oblate
                                 spheroid corresponding to  an ellipse of revolution about its short axis. 
                                 He calculated  the ellipticity $\epsilon=a/b-1$, where $a$ and $b$  are the major and
                                 minor axis of the ellipse,
                                  by the requirement  that two  channels of water  to the center of the earth, one
                                   from the pole and  another from the equator
                                   would be in pressure equilibrium at the center. Remarkably, 
                                   in his calculation Newton also  took into   account the variation
                                   of the gravitational force inside the earth due to the  shape distortion  which he discussed
                                   in Prop. 91,  Cor. 3, Book 1 (For a  modern discussion see (Whiteside 1974, 225-226) and (Chandrasekhar
                                   1995, 313-317) . 
                                   Newton  obtained 
                                   for  the ellipticity,  $\epsilon=(5.04 /4) \delta $, where $\delta $, the ratio
                                   of centrifugal acceleration to the acceleration of gravity $g$, is  $\delta=(4\pi^2 r_e /(g T^2)$,
                                   $r_e$ is the mean radius of the earth and  $T$ is the period of rotation ( one siderial day).
                                   This gives  $\delta=1/289$, and Newton  found that $\epsilon=1/229$ and
                                   announced  that the distance to the center of the earth  at the equator  exceeds the
                                   value at the poles  by $\epsilon r_e$= 17 miles. The present observed value is 13 miles,
                                   because the  actual density
                                   of the earth is not homogeneous.       
                                    A similar calculation was carried out
                                   by Huygens  who included, however, only the effect of the centrifugal forces, because he
                                   did not accept Newton's principle of universal gravitation.  Hence, Huygens obtained
                                   $\epsilon=(1/2)\delta = 1/578$.  Newton's result was first  derived by
                                   Clairaut, who showed that the relation $\epsilon=(5/4)\delta$ is correct to first order
                                   in $\epsilon$ (Todhunter 1962, 204).
                                     }.
This prediction   was  contrary to the observations  of  the  Italian astronomer Gian Domenico Cassini, who had joined
the French Academy of Sciences, and his son Jacques Cassini. They had
obtained faulty geodetic measurements, indicating 
 that the earth is a prolate spheroid \fo{\no   For example,  Jacques Cassini found that the length of a degree of longitude  in the parallel of St. Malo, France,
                                     is 36,670 {\it toises},  but  on the supposition of a spherical  earth it should be 37,707
                                     toises (Todhunter 1962, 111) ( The length of a  {toise} can  be obtained
                                     from Newton's remark  in Prop. 19, book 3,  that 367,196 {\it London} feet,  the mean
                                     radius of the earth obtained by a Mr.  Norwood,  is  equal 57,300 {\it Parisian} toises).
                                      }.                               
To resolve this conflict,  Maupertuis together with the French mathematician Alexis-Claude 
 Clairaut  led a scientific  expedition commissioned by the French Academy of Sciences that  left for Lapland on April 20, 1736
 to measure the length of a degree of the meridian at that latitude, in order  to
compare it with the corresponding  length at the latitude of Paris.  Mauepertuis 
became famous for confirming Newton's prediction, and  Voltaire called  him the 
 ``aplatisseur du monde et de Cassini",  remarking sarcastically  that 
	                           \beq
	                           Vous avez confirm\'{e} dans des lieux pleins d'http://arxiv.org/help/ ennui\\
	                           Ce que Newton connut sans sortir de chez lui 	                        	                           
	                           \fo{\no
                                   You have confirmed in these difficult locations 
                                   what Mr. Newton knew without leaving his  home.(Florian 1934, 664)
                                   }
	                           \enq
	                          Another expedition,  headed by La Condamine,  Bouguer and Godin,  also members of the French Academy 
of Sciences,  went about  a year earlier to Peru to measure a
corresponding  arc of the meridian near the
equator. But they ran into considerable difficulties and delays due to 
personal animosities between the leaders of the expedition, and only
 ten years later,  were they  able to
report their results which  were  consistent with the conclusions of  Maupertuis' expedition (Todhunter 1962). 
Subsequently,  the problem of evaluating theoretically  the shape of the earth became the subject of intense efforts
by  Continental mathematicians who studied  Newton's {\it Principia}, and its difficulty
spurred  major advances in mathematical physics.
	                     
 In his {\it Lettres philosophiques}, Voltaire 
reported these controversies with his characteristic wit,
\beq
For your Cartesians everything is moved by an impulsion you don't really understand,  while  for Mr. Newton
it is by gravitation,  the cause of which is hardly better known.  In Paris you see the earth shaped
like a melon, in London it is flattened on two sides (Voltaire 1734)
\enq
The contrast between the methods of  Descartes and Newton were neatly  contrasted 
by Bernard le Bovier  Fontenelle,  the secretary of the French Academy of Sciences, who wrote in his {\it Eloge de Newton},
\beq
Descartes proceeds from what he clearly understands to find the cause of what he sees,  whereas
Newton proceeds from what he sees to find the cause, whether it be clear or obscure (Fontenelle  1728)
\enq
Newton, however,  left open the question of the origin of the gravitational force. During a correspondence
with the Reverend Richard Bentley,  he made his reservations  clear,
\beq
It is  unconcievable that inanimate brute matter should ( without the mediation of something else
which is not material) operate and affect other matter without mutual contact; as it must if gravitation
in the sense of Epicurus be essential and inherent in it ... That gravity ... may act at a distance  through
a vacuum ... is to me so great an absurdity that I believe no man who has in philosophical matters any
competent faculty of thinking can ever fall into it. Gravity must be caused by an agent acting
constantly according to certain laws, but whether this agent is material or inmaterial is a question
I have left to the consideration of my readers  (Westfall 1995, 505).
\enq
This view  led to accusations by some readers of the {\it Principia}
 that Newton had left  the physics out of his mathematics.

 Kepler had shown that the planets travel  along elliptical orbits with the sun located at one of the foci
\fo{\no This subtlety was not always appreciated by the general public. For example,
                            the  Bank of England  issued
                             a two  pound  note, now retracted, 
                              showing incorrectly a figure from Newton's {\it  Principia}, 
                             with the  sun  at the center of the ellipse.},
moving with  a non-uniform velocity  satisfying his  second law or area law\fo{\no
 In the  {\it Principia}, Book I, Prop. 1,  Newton gave the following 
formulation of the area law:
\beq
The areas which bodies made to move in orbits describe by radii drawn from an unmoving center
of forces lie in unmoving planes and are proportional to the times (Cohen 1999).
\enq}.
To account for such an orbit, Newton had to extend   
the rigorous geometrical methods developed by  Greek mathematicians 
to encompass the limit of ratios and sums of  { \it vanishing} quantities (Nauenberg 1998).  In
the { \it Principia} such quantities  were represented by  lines and  arcs of curves of arbitrarily
small length, a procedure  that  had been introduced  by Appollonius, and applied
by Ptolemy for calculations in his geocentric model  of celestial motion, and 
by Archimedes for calculations of 
lengths and areas encompassed by curves. In the  17-century,  this procedure was developed further
 by several mathematicians  including in particular Ren\`{e}  Descartes, whose work  Newton had 
studied  carefully \fo{\no    Newton studied   Frans van Schooten's  second edition (1659) of his
                                 translation of Descartes {\it Geometrie} from French into Latin, with appended tracts
                                 by Hudde, Heurat and de Witt.  This translation was crucial to Newton's education
                                 because he could not read French.} (Whiteside 1967).
                                   Since motion occurs  with the passage of time,  it was necessary
for Newton to express  time as  a geometrical variable, but this  was  a  major stumbling block (Nauenberg 1994a). It was  only  
after a lengthy correspondence  with Robert Hooke (Turnbull 1960, 297-314) (Nauenberg 1994b), that  Newton 
was able to give a proof of the validity of  Kepler's area law for any central force  (Nauenberg 2003). Newton recalled  that 
                                   \beq
                                     In the year 1679 in answer to a letter from Dr. Hook ... I found now that
                                    whatsoever was the law of the forces which kept the Planets in their Orbs,
                                    the area described by a radius from them to the Sun would be proportional to 
                                     the times in which they were described. And by the help of these two propositions
                                     I found that their Orbs would be such ellipses as Kepler had described... (Lohne 1960)
                                                                         \enq
 Thus, Newton was able to  geometrize the  passage of time  by the  
change of an area -  a concept without which writing the
 {\it  Principia} would not have been possible. He   
emphasized its importance by starting the {\it  Principia} with a  mathematical proof of the
generalization of Kepler's second law described  in Prop. 1, Book 1.
 (Brackenridge 1995) (Nauenberg 2003).
  
 The style  of the {\it  Principia}
 followed the mathematical format  of  the  {\it Horologium Oscillatorioum}  by Christiaan Huygens, who was the
 most prominent  scientist in the Continent during  the later part of the 17-th century (Huygens 1673)
  (Nauenberg 1998).  
  In 1673, when Newton received a copy of Huygens' book from Henry Oldenburg,  he promptly
  responded that
  \beq
  I have viewed it with great satisfaction finding it full of very subtile and useful speculations very
  worthy of the Author.  I am glad, we are to expect another discourse of the { \it Vis Centrifuga} [centrifugal
  force ] which speculation may prove of good  use in natural Philosophy and Astronomy, as well
  a Mechanicks. (Huygens 1897, 325)
  \enq
 
 In the preface of his biography, {\it A view of Sir Isaac Newton's Philosophy},
 Henry Pemberton, who was the editor of the the third edition of the {\it Principia},  wrote that
  \beq
  Sir Isaac Newton has several times particularly recommend to me Huygens' style and manner.
  He thought him the most elegant of any mathematical writer of modern times, and the
  most  just imitator of the ancients. Of their taste, and form of demonstration Sir Isaac always
  professed himself a great admirer. I have heard him even censure himself for not following them
  yet more closely than he did; and speak with regret of his mistake at the beginning of his mathematical
  studies , in applying himself to the works of Des Cartes  and other algebraic writers, before he had
  considered the elements of Euclide with that attention, which so excellent a writer deserves (Pemberton 1728).
  \enq
 In turn Huygens greatly admired Newton's work, and in the summer of  1689 he came to
 England  to meet Newton and discuss  with him the current theories of  gravitation.  
 Like Leibniz,   Huygens did not accept  Newton's concept of an action at a distance which was
 regarded as an occult force by  followers of Descartes vortex theory of gravitation, but he accepted
 the inverse square dependence on distance of the gravitational force.
 
 In 1689  the British mathematician David Gregory visited Newton in Cambridge,  and  reported  that 
\beq
 I saw a manuscript [written]
before the year 1669 ( the year when its author  Mr. Newton was made Lucasian Professor of Mathematics) where all the foundations of his philosophy are laid: namely the gravity of the Moon
to the Earth, and of the planets to the Sun. And in fact all these even then are subject 
to calculation (Herivel 1965, 192).
\enq

This  manuscript, which Newton never published,  revealed that already sixteen years before
 writing the {\it Principia},
 Newton had carried  out the ``moon test", later described in Book 3,  Prop. 4,  where  he compared the gravitational
force of the earth on the moon to the force of gravity on an object on the surface of the earth (Herivel 1965, 192-198)(see Appendix).  
In order to make this
comparison, however, Newton had to assume  that the gravitational force varied inversely proportional to the square of the distance from the center of the earth to any distance above its  surface. Previously, he had  deduced the inverse square  dependence of this force  from  planetary motion ( see footnote 2), where the distance between the planets and the sun is very large compared to their sizes,  and  
it was reasonable to treat these  celestial bodies  as  point masses.  But to assume that this radial dependence was still valid for  much shorter distances,  and in particular down to the surface of a 
planet had to be justified.  Apparently it was only after Newton  already had started writing the {\it Principia}, 
that he was able   to provide such a justification, by assuming that the gravitational
attractive force due a  finite size body   can be compounded by adding the contribution  of each 
of its  elements. In Prop. 71, Book 1 of the {\it  Principia} he  gave  a remarkable proof 
that the gravitational force of  a
spherical distribution of mass acts at any distance from its surface  as if the total mass is concentrated 
at its centre (Chandrasekhar 1995, 269-272).  Furthermore, in Prop. 91, Book 1,  he considered  also  the force acting along the axis of any solid
of revolution, and in Cor. 2  he applied the result  to evaluate  the  special case of an oblate ellipsoid  which he needed to
determine the eccentricity of the earth due to its daily rotation ( see footnote 6).

In Book 3 of the {\it Principia}, Newton applied his mathematical theory of orbital
dynamics to planetary and lunar motion and to the motion of comets in order 
to provide evidence for the  universal law of gravitation - that the attractive gravitational force between two
bodies is  proportional
to the product of their  masses and inversely proportional to the square of the distance
between them \fo{\no In `Rules of Reasoning in Philosophy', { \it Principia}, Book 3, in Rule 3
                   Newton concluded :

                   \beq 
                   Lastly, if it universally appears, by experiments and astronomical observations, that all
                   bodies about the earth gravitate towards the earth, and that in proportion to the quantity
                   of matter [mass] which they severally contain; that the moon likewise, according to the 
                   quantity of its  matter, gravitates towards the earth; that on the other hand, our sea gravitates
                   towards the moon; and all the planets one towards another; and the comets in like manner
                   towards the sun; we must in consequence of this rule, universally allow that all 
                   bodies whatsoever are endowed with a principle of mutual gravitation.                    
                   \enq
                   Compare Newton's formulation of universal gravitation
                   with the  earlier one of  Robert Hooke, who  wrote,
                   \beq
                   That all Celestial Bodies whatsoever, have an attraction or gravitating power towards their
                   own Centers, whereby they attract not only their own parts, and keep them from flying from them, as
                   we may observe the Earth to do, but that they do also attract all the other Celestial  Bodies that are
                   within the sphere of their activity; and consequently that not only the Sun and Moon have an influence
                   upon the body and motion of the Earth, and the Earth upon them, but that Mercury, also Venus, Mars  
                   Saturn and Jupiter by their attractive powers, have a considerable influence upon its motion as in the
                   same manner the corresponding attractive  power of the Earth hath a considerable influence upon every
                   one of their motions  also (Hooke 1674) (Nauenberg 1994a).
                   \enq} .
He   persuaded the Royal Astronomer, John  Flamsteed,  to provide him 
with the best available  observational data at the time for the periods and major axis of the planets,  
and for the Jovian and Saturnian satellites. Then he showed 
 that  these observations  were  in good  agreement with Kepler's third law, Book 3, Phenomenon 2 - 4,  which 
 for circular motion  he had considered some 20 years earlier  as formulated  in  Cor.6 of Prop.4, Book 1, (see Appendix)
\beq 
 If the
periodic times are as the three half powers of the radii, the centripetal force will
be inversely as the squares of the radii.
\enq

In Prop.15,  Book 3,  he extended this proof to elliptical motion,  applying  Cor.1 of Prop.45,
 Book I, to show that the near immobility of the aphelia of the planets, Book 3, Prop. 14, implied that the
  gravitational  force between the planets and the sun satisfied the inverse square law. 
  This was  Newton's best proof for the inverse square law, because 
  he had shown  that the smallest    deviation from this law would give rise to
  a precession of  the planetary  aphelia  which over the centuries would have accumulated to give an
  observable effect. 
 
 Newton was aware,  however, that astronomical observations had shown
that there were deviations from Kepler's laws  in the motion of the planets  and
the moon. In the preface to the  {\it Principia}, he wrote:
\beq
But after I began to work on the inequalities of the motion
of the moon, and ...the motions of several bodies with
respect to one another ...I thought that publication should be put
to another time  so that I might investigate these other things...
\enq  
Remarkably, a large part of these  investigations apparently took place during the time
that Newton was composing  his book,  when he developed 
 methods  to calculate the perturbation of the solar gravitational force on the lunar motion
 around the earth,
 and the effects  due to the interplanetary
gravitational forces on the motion of the planets around the sun.  In Prop. 45
he presented his simplest  perturbation approximation for the lunar orbit 
by  assuming that  it   was a Keplerian  elliptic orbit, but  with its major axis rotating
uniformly.  In  characteristic fashion,  first he solved the problem of 
obtaining the exact  law of force which would give rise to such an orbit,
and found that this force was a linear combination of inverse square and inverse
cube forces . Then he  determined the  rotation rate  of the lunar apse by 
considering the effect  of the
component of the solar  gravitational force along the earth-moon radial distance averaged
over a period. But  this approximation gave a precession of the major axis of the lunar ellipse of only half the
observed  rate.
In the first two editions of the {\it  Principia} Newton was somewhat ambivalent about this large discrepancy,
and only in the third edition,  which appeared in 1726,   did he add a remark,
in Corollary 2 of Prop. 45, that `` the apse of the Moon is about twice as swift "
as the value that he had calculated.
This discrepancy  became one the first major challenges for the mathematicians and
astronomers who studied  the {\it Principia}, and it took another 20 years before a 
solution to this problem was first found by  Clairaut and the French mathematician
Jean le Rond d'Alembert.

\section*{ The Reception of the  {\it Principia} by the Mathematicians in the Europe }
  
  When Continental mathematicians and astronomers, 
  primarily from Holland, Germany, Switzerland, and France,  first read the  {\it Principia}, 
 they had  some difficulties understanding Newton's novel 
 mathematical concepts, with its   combination of  geometrical quantities in the tradition
 of Greek mathematics and  his concept of  limits  of ratios and sums of infinitesimals - quantities
 which become vanishingly small (Nauenberg 2010). After introducing  three ``Laws of Motion'', Newton
 presented  ten mathematical  ``Lemmas"  on  his geometrical differential method of  `` first and last 
 ratios". These lemmas  constitute  the basis for  his calculus,  and he referred 
 to them in the proof of his propositions. Except for Lemma 2 in Book 2 of the {\it Principia}, Newton did not explain his analytic
 differential calculus
 in much detail,  and
European  mathematicians, who already had been introduced to an equivalent calculus \fo{\no
In 1696 the Marquis de l' Hospital published {\it Analyse des infiniment petits},  based on lectures about
the calculus of Leibniz, 
given to him by Johann Bernoulli who he had hired as  his  private  tutor in mathematics.}
 by 
the German philosopher and mathematician, Gottfried Wilhelm Leibniz,  first 
had to  translate  Newton's mathematical language into Leibniz's  language before
they could  make further progress. Indeed, Leibniz was  the first to express Newton's 
formalism for orbital motion in the form of a differential equation based on his calculus (Nauenberg  2010).
Leibniz claimed to have achieved  his results having only read a review of Newton's {\it Principia},
but  in 1969  E.A. Fellman (1973) obtained  a copy  of  Newton's work which contained  abundant
Marginalia by Leibniz,  indicating that he had carefully studied  the text  before undertaking his
own work. Moreover, recently discovered manuscripts show  Leibniz preparatory work for his
1689 essay {\it Tentamen de motuum coelestium causis} based on a reading of the {\it Principia}
(Aiton 1995), 10) (Bertoloni Meli, 1991).  But  Leibniz obtained  a differential
equation of motion for celestial objects that was remarkably original \fo {\no Applying Prop. 1 in Book 1 of the {\it Principia}, Leibniz derived an expression for the
second order differential $ddr$ for  the radial distance.  This led him to a genuine discovery which is not
found in the {\it Principia}: that this differential
is proportional  to an effective centrifugal force minus the central attractive  force $f(r)$. In modern notation Leibniz's result 
corresponds to the equation $ d^2r/dt^2= h^2/r^3 - f(r)$, where $h=r^2 d\theta/dt $ is a constant corresponding
to the angular momentum. For the case that the orbit is an ellipse he found 
that $f(r)=\mu/r^2$, where $\mu$ is the strength of the gravitational interaction (Aiton 1960) (Aiton1995) (Bertoloni Meli 1991) (Guicciardini 1999). Leibniz, however, assumed  without justification that $h=\mu$= latus rectum of the ellipse, 
and he incorrectly  attributed Kepler's area law to a property of celestial vortices which leads to a physically inconsistent interpretation
of his equation.} 
Another  mathematician who  applied  Leibniz's version of the differential  calculus  was   Jacob Hermann, a member
of a group around  Jacob and Johann Bernoulli,  two of  Europe's  leading mathematicians
who had formed a school in Basel (Guicciardini 1999).  Expressing Prop. 1 and Prop. 2 in Book 1 of Newton's
{\it Principia} in the language of this calculus, he obtained a differential  equation for the motion
of a body under the action of central force. Then he gave a proof that conic sections where the only solutions for
the case that the central force  varied inversely with the square of the distance from the center of force (Herman 1710), 
(Nauenberg 2010)
 This was an important result, because in the first edition of the
{\it Principia},  in Cor. 1 to Prop. 13, Newton had asserted, without proof,  that conic sections curves were the
{\it unique} solutions  to orbital motion  under inverse square forces.  Johann Bernoulli  
criticized  Hermann's solution  for being incomplete( Hermann had left out a constant of the motion), and then  
derived the elliptic orbit by solving,  via  a suitable transformation of variables, 
 the   general integral 
for orbital motion in a central field force given in Prop. 41 Book 1 of the
{\it  Principia},  for the special case of an inverse square force (Bernoulli 1710), (Nauenberg 2010) 
Remarkably, Newton did not include
this fundamental solution in the  {\it  Principia},  giving rise to  a  gap  that  has  caused considerable
confusion in the literature that remains  up to the present time. Instead, Newton gave as
an example the orbit  for an inverse  cube force \fo {\no Prop. 41, Cor. 3} \footnote{
 In Prop. 11-13  Newton gave  a proof that 
if the orbit for a central force is a conic section, then the force varies inversely
as the square of the radial distance.  Johann Bernoulli  criticized the incompleteness of
Cor. 1 of Prop. 13, Book 1, where Newton claimed to give  a proof
to the solution of   the {\it  inverse} problem: given the gravitational force to show that
the resulting orbit is a conic section\fo {\no
In the 1980's,  Bernoulli's  criticism of Cor. 1 to Prop. 11-13  was 
                                       revived  by  Robert Weinstock,
                                      in `Dismantling a centuries-old myth: Newton's Principia and inverse
                                      square orbits', {\it  American Journal of Physics} 50 (1982) 610-617.  
                                      Weinstock's arguments are dismantled in  (Nauenberg 1994c) . }                                                                         }. 

 Bernoulli  also communicated to Newton an error
he had found in Prop. 10 of Book 2.  In both cases Newton made corrections in the
next edition of the Principia (1713) without, however,  acknowledging Bernoulli's important
contributions (Guicciardini 1999). Some British mathematicians like
David Gregory were able to contact Newton, and get help from him to overcome obstacles
in understanding the Principia, but this appears not to have been possible for 
Continental  mathematicians. 

After Leibniz, the first  Continental mathematician who  undertook the reformulation  of  Newton's mathematical 
concepts into the language of Leibniz's calculus, was Pierre Varignon (Aiton 1960, 1955) (Bertoloni-Meli 1991) (Guicciardini 1999) .
Varignon introduced an alternative expression for a  central force in terms of the 
 curvature of the orbit\fo{\no Varignon (1701) called the radius of curvature  `le rayon de D\'{e}velopp\'{e}',  and
 obtained his  expression for the central force by recognizing
 that a small arc of the orbit corresponds to that  of a circle with this radius, called the  {\it osculating} circle
 by Leibniz but originating  in the work of Huygens  (Kline 1972) (Nauenberg 1996),
 },  without being aware that  Newton's earliest  understanding  of non-circular orbital motion was also based on curvature (Nauenberg 1994). 
 In a cryptic note written in his 1664  Waste book,  Newton remarked  that
\beq
If the body b moved in an ellipse,  then its force in each point (if  its motion in that point
be given) [can] be found by a tangent circle of equal crookedness with that point of the
ellipse (Herivel 1965, 130)
\enq
Here the word ``crookedness"  refers  to  curvature which is measured locally by  the
inverse radius  $\rho$ of the
tangent or osculating circle (as it was named later by Leibniz) at any  point on  an ellipse. Curvature
was also a mathematical concept that had been introduced earlier by Huygens 
in his {\it Horologium Oscillatorum} (Huygens 1673) , (Nauenberg 1996)
Evidently, Newton was aware that
the  central force or acceleration $a$ for non-uniform orbital motion
can be obtained from  a generalization
of the relation for  uniform circular  motion, $ a_c=v^2/\rho$, where $v$ is the velocity,
which he, and independently Huygens, had obtained earlier (see Appendix). Then
$a=a_c/cos(\alpha )$ where $\alpha$ is the angle between the  direction of the central force
and that  of the radius of curvature. The problem, however, is that the motion
or velocity $v$,  which is a variable along the orbit for non-circular motion because then
there is a tangential component of the central force, had to  be known (Brackenridge 1995) , {Brackenridge 2002 ).
But 15 years later,  Newton found a proof for the area law,
which  implies that  for any  central force, the area swept by the radial line per unit time, 
  $(1/2)vr cos(\alpha)$ is a constant (proportional
to the conserved angular momentum) and $r$ is the radial distance. By substituting this expression
for $v$,  Newton had  an
explicitly expression for  the central
acceleration  $a \propto 1/ \rho r^2 cos^3(\alpha)$.   Indeed, for conical sections,
the quantity $\rho cos^3(\alpha)$ is a constant (the semi-latus rectum of an ellipse) which provided Newton with a succint
proof that  for such orbits the force depends inversely  on the square of the
radial distance (Nauenberg 1994a). This relation was also found  by 
Abraham DeMoivre\fo{\no  \beq  After having found this theorem, I showed it to M. Newton and I was proud to believe that it would have appeared 
new to him, but M. Newton had arrived at it before me;  he showed this theorem  to me among his papers that he is
preparing for a second edition of his {\it  Principia Mathematica } ...
\enq} 
 (Guicciardini 1999, 226), and applied by John Keill and  Roger Cotes who were  members of the school of
 British mathematicians \fo{\no For a brief history of this important development see
 (Whiteside 1974, 548-549)}.
 In the first edition of the {\it Principia},
however, the curvature expression for the force does not appear explicitly,  although Newton applied
it  in a few instance without any explanation \fo {\no
 Prop. 15, Book 2, and Prop. 26-29 Book 3.}, while in the second edition curvature  is discussed
in a new Lemma, Lemma 11, and the curvature measure for force is derived
as  corollaries to Prop. 6.  Subsequently, Newton  applied it  to obtain 
  ``another solution"  to  the fundamendal  problem formulated in   Prop. 11, Book 1,
 \beq 
  Let a body revolve in an ellipse,
   it is required to find the law
  of the centripetal [central] force  tending towards a focus of the ellipse.
 \enq
 According to Newton's recollections, as  told  to DeMoivre in 1727,
                              \beq
                              In 1684 Dr. Halley came to visit him at Cambridge and after they had been
                              some time together, the Dr. asked him what the thought the Curve would
                              be that would be described  by the Planets supposing the force  of attraction
                              towards the Sun to be reciprocal to the square of their distance from it. Sir
                              Isaac  replied immediately that it would be an {\it Ellipsis}, the Dr. struck with
                              joy and amazement asked him how he knew it, why said he, I have calculated
                              it, whereupon Dr. Halley asked him for his calculation without delay. Sir Isaac
                              looked among his papers but could not find it, but he promised him to renew it,
                              and then to send it to him. (Westfall 1995 ).
                               \enq
 However, the solution which Newton eventually sent to Edmund Halley in a manuscript entitled
 {\it  De Motu} (Whiteside 1974), and that  three years later he presented  in the same form
  in Prop. 11 of  the {\it Principia},
 treated  instead  the  inverse to the  problem posed by Halley,  namely  given that the orbit is an ellipse, to prove that the
 central force obeys the  inverse square law, or as Newton formulated in 1687, 
 \beq
 Let a body revolve in an ellipse; it is required to find the law of the centripetal force tending
 towards a focus of the ellipse (Cohen 1999, 462)
 \enq

Relations with  Bernoulli and his school were further aggravated when the notorious 
priority dispute on the invention of the calculus erupted between Newton and Leibniz
in 1711. By the 1740's, serious reservations arouse regarding the general
validity of the inverse square law for gravitational force because of the failure
of Newton?s approximation of the solar perturbation to account for the rate of
precession of the lunar apside. One of the first to question on this ground the validity
of this law was the great mathematician Leonhard Euler.
He remarked that,
\beq
having first supposed that the force acting on the Moon from
both the Earth and the Sun are perfectly proportional reciprocal to the squares
of the distances, I have always found the motion of the apogee to be almost
two times slower than the observations make it: and although several small terms that
I have been obliged to neglect  in the calculation may be able to accelerate the motion of the 
apogee, I have ascertained after several investigations that they would be far from sufficient
to make up for this lack, and that it is absolutely necessary that the forces  by which the
Moon is at present solicited are a little different from the ones I supposed. (Waff 1995, 37).
\enq
He concluded that 
 \beq
  all these reason joined together appear therefore to prove 
 invincibly that the centripetal force in the Heavens do not follow exactly the law 
 established by Newton (Waff 1995, 37).
 \enq
Clairaut  had reached similar conclusions
and was delighted to find that he was in agreement with Euler. He had also found 
\beq
that the period of the apogee [i.e. the time it takes for the lunar apogee to return
to the same point in the heavens] that follows from the attraction reciprocally proportional
to the squares of the distances, would be about 19 years, instead of a little  less than
9 years which it is in fact (Waff 1995, 39)
\enq
a result that Newton had mentioned earlier in Prop. 45, Book 1.
To account for this discrepancy, Clairaut proposed that an additional force was also in effect 
which varied with distance inversely as the fourth power, possible due to Cartesian
vortices.  Actually,  suggestions  for  possible correction to the inverse square
gravitational law  had been considered by Newton in Query 31 of his {\it Opticks} , but 
he did not want to publicize them.   Another mathematician, 
Jean le Rond d' Alembert,  arrived at the same  discrepancy  for the motion of the 
lunar apogee, but  in contrast to Euler and Clairaut,  he
did not questioned the mathematical
form of Newton's  gravitational law because of its successes in describing 
other inequalities of  the lunar motion.  Ultimately, the French Academy of Sciences propose
a prize for the solution of this problem, and  in 1749  Clairaut  finally obtained a solution
without altering the inverse square force, by considering higher order contributions to the solar perturbation,
followed by d' Alembert  with a more careful analysis which gave the same result.
\fo {\no The title that Clairaut chose for his winning essay was `Theory of the Moon Deduced
from the Single Principle of Attraction Reciprocally Proportional to the Squares of
the Distances'} (Waff 1995). Previously, similar solution  had been  obtained by Newton,
but it contained some errors (Nauenberg 2001a),  and in a  Scholium to Prop. 35, Book 3,  inserted only  in the  first edition
of the {\it Principia},  he declared that
\beq
...These computations, however, excessively complicated and clogged with approximations as they
are, and  insufficiently  accurate we have not seen fit to set out .
\enq
The details of Newton's  computations  remained unknown  until  1872 when they were  found among his papers
in the Portsmouth Collection (Whiteside 1974, 508-538) (Nauenberg 2000) (Nauenberg 2001a)

The importance of Clairaut's  result can hardly be overestimated. In admiration Euler
declared in a letter to Clairaut that 
\beq
. . . the more I consider this happily discovery,
the more important it seems to me. For it is very certain that it is only since this
 discovery that one can regard the law of attraction reciprocally proportional to
the squares of the distance as solidly established, and on this depends the entire
theory of astronomy (Waff 1995, 46)
\enq
In 1748 the French academy of sciences chose for its prize contest  a theory that would explain the
inequalities in the motion of Jupiter and Saturn due to their mutual gravitational
interaction, which Newton had considered only semi-quantitatively in Prop. 13, Book 3 \fo{\no
From the action of  Jupiter  upon Saturn ``...arises a perturbation of the orbit of Saturn at every conjuction of
this planet with Jupiter , so sensible, that astronomers have been at a loss concerning  it" ( Cohen 1999, 818).}   
This  problem was much  more difficult  than the lunar case,  and  Euler was the first
to deal with it (Euler  1769), and  now he declared that
\beq
... because  Clairaut has made the important  discovery that the movement of the apogee of the Moon
is perfectly in accord with the Newtonian hypotheses ..., there no longer remains the least doubt about
its proportions... One can now maintain boldly that the two planets Jupiter and Saturn attract each other
mutually in the inverse ratio of the squares of their distance, and that all the irregularities that can be discovered
in their movement are infallibly caused by their mutual action... and if the calculations that one claims to have
drawn from the theory are not found to be in good agreement  with the observations, one will always be
justified to doubting the correctness of the calculations, rather than the truth of the theory (Waff 1995, 46) 
\enq

After missing an expected lunar eclipse, Tycho Brahe had discovered  a bi-montly variation in the lunar speed, 
and  Newton was able to account for this variation as an effect of the solar gravitational force.
In Prop. 28, Book 3, Newton introduced a novel frame of reference where
the earth is fixed at the center of  a  rotating frame  with the period of one year.  In this 
frame the sun stands still  when  
 the eccentricity of the earth-sun orbit is neglected. Then  taking into account the solar gravitational force,
Newton found an approximate periodic orbit of the moon which accounted for the periodic  of the
variation discovered by Brahe.  In Prop. 29, Book 3,  appropriately 
entitled  `To find the variation of the moon',
he calculated  the amplitude of this variation, and found it in very good agreement with Brahe's  observation.
In his review of Newton's work on lunar theory  the great French mathematician and
astronomer, Pierre-Simon Laplace,  singled  out this result,
and remarked admiringly at Newton's insightful approximations,
\beq
Such hypothesis in calculations ... are permitted to inventors during
such difficult researches  \fo{\no
                                        Ces hypoth\`{e}ses de calcul...  sont permises aus inventeurs  
                                        dans des reserches aussi difficiles (Laplace 1825, 391)
                                          }
\enq

\section*{Reception of Newton's gravitational  theory for planetary and lunar motion}
Inspired by Newton's work,  
 Leohnard  Euler introduced  his  rotating frame to calculate the solar  perturbation to  the lunar motion (Euler 1772). Likewise, in  1836 
 this frame   was 
considered  also by Gustaf Carl Jacobi,  who gave a proof for the
existence of a constant of the motion in what became known as the
restricted three body problem.  Later, 
the American astronomer  George Hill  also obtained   periodic  solutions in this rotating frame 
(Hill 1783), and his work was extended by
Henri Poincar\'{e}, which led him eventually to his profound discovery
of  {\it  chaotic} orbital motion in Newtonian dynamics  (Poincare 1892 ), (Barrow-Green 1991 ), (Nauenberg  2003b ). 

In twenty two  corollaries to Proposition 66, Book 1, Newton  described entirely in prose   his perturbations methods,    
but his detailed calculations remained unpublished (Nauenberg  2000) (Nauenberg 2001a). 
Here Newton considered  gravitational  perturbations to the elliptical motion of a
planet around the sun or  the moon around the earth as  a sequence of 
impulses, equally spaced in time,  which instantaneously alters  the velocity of the celestial body in its orbit without, however, changing its position when these impulses occur.  In  Prop. 17, Book 1,
Newton had shown how the orbital parameters - the eccentricity, and the magnitude and direction
of the principal axis  of the ellipse - can be  determined given the velocity and position at a
given time (initial conditions). Hence,  these  impulses 
lead  to periodic  changes in the orbital parameters which are determined by the discontinuous change
in velocity after the impulse has taken place. In corollaries 3 and 4 of Proposition 17, Newton gave a succint description of his  method of variation  of orbital parameters .  These corollaries were
added to  later drafts of the  {\it Principia} \fo {\no In the initial revisions of the early manucript for the {\it Principia},   Prop. 17 
                                contained only Corollaries 1 and 2 (Whiteside 1974, 160--161)}.
  indicating that Newton had developed this method
during the period when he was writing his book.
  In the limit that the time interval between impulses
 is made vanishingly small, Newton's perturbation methods  corresponds to the method
 of variational parameters developed much later by Euler 
 \fo{ \no  Starting with the equations
of motion as second order differential
equations in polar coordinates, Euler assumes that the solution for the
orbit is described by an ellipse with time varying orbital parameters
$p$,$e$ and $\omega$, where $p$ is the semilatus rectum of the ellipse,
$e$ is the eccentricity, and $\omega$ is the angle of the major axis.
Then he obtained  first order differential equations for $e$ and $\omega$
by imposing two constraints: that $p=h^2/\mu $, 
where $h$ is the angular momentuma, and that
$E=\mu (e^2-1)/2p$  where E is the time varying Kepler energy 
of the orbit. In modern notation $\mu = GM $ where $M$ is 
the sum of the mass of the earth, and the moon and $G$ is 
Newton's gravitational constant (Euler 1769). It can be readily shown
that Euler's constraints lead to the same definition of the ellipse 
described  geometrically by Newton in the Portsmouth manuscript ( Nauenberg 2000) (Nauenberg 2001a).},
Joseph Louis Lagrange and Pierre-Simon Laplace \fo {\no    Laplace
                           obtained the differential equations for the time
                           dependence of the orbital parameters 
                           by evaluating the time derivate  of  the 
                           vector  $\vec {f}=\vec {v} \times \vec {h} - \mu 
                           \vec {r}/r$, where $\vec {f}$ is a vector
                           along the major axis of the ellipse with
                           magnitude $f=\mu e$.  The construction of this
                           vector was first  given in geometrical form  
                           by Newton in Book 1, Prop. 17, and  in
                           analytic form by Jacob Hermann and Johann Bernoulli (Bernoulli,1710) (Nauenberg 2010).
                           Laplace's derivation(Laplace 1822, 357-390) of the
                           variation of orbital parameter is in effect  
                           the analytic equivalent of Newton's geometrical
                           approach in the Portsmouth manuscript (Nauenberg 2000) (Nauenberg 2001a)}.
                           Now  this method  is  usually credit to them.   

 Unpublished manuscript  in the Porstmouth collection of Newton's papers, first examined 
 in 1872 by a syndicate
 appointed by the University of Cambridge (Brackanbridge 1999)  reveal that Newton had 
 intended to include a more detailed  description of
  his perturbation methods in the {\it Principia}, but  neither the propositions and
  lemmas in these manuscripts nor the 
  resulting equations, which in effect are non-linear coupled differential equations for the
 orbital parameters,  appeared  in any of its  three editions (Nauenberg 2000) Nauenberg 2001a). 
 But some of his results for the inequalities of the lunar motion
 appeared in a lengthy Scholium  after  Prop. 35, Book 3,  which includes numerical results
 obtained by approximate solutions to his equations.
 In this Scholium, for example,  Newton stated that

\beq
By the same theory of gravity, the moon's apogee goes forwards at the
greatest rate when it is either in conjunction with or in opposition to
the sun, but in its quadratures with the sun it goes backwards; and 
the eccentricity comes, in the former case to its greatest quantity;
in the latter to its least by Cor. 7,8 and 9 , Prop. 66, Book 1.
And those inequalities by the Corollaries we have named, are very great,
and generate the principle which I call the semiannual equation of
the apogee; and this semiannual equation in its greatest quantity
comes to about $12^o18'$, as nearly as I could determine from the
phenomena \fo {\no In Cor. 7 and 8, Prop. 66, Newton gave a qualitative
explanation for this  motion
of the moon's apogee due to the perturbation of the sun, 
stating  it was based on results given in Book 1, Prop. 45, Cor. 1.
However, these results were obtained for the case of 
radial forces only, and are therefore strictly
not applicable to the solar perturbation  which is
not a purely radial force with respect to the earth as a center, 
and which depends also on the angle $\psi$ .
According to the differential equation for the  motion
of the lunar apogee which appears in the Portsmouth manuscript,
,his rate depends
on the relative angle between the moon's apogee $\omega$ and
the longitude $\theta$ of the sun, where $\omega -\theta = \psi -\phi$. It reaches a maximum
value when $ \omega - \theta= n\pi$ where $n$ is an integer.
and a minimum when $n$ is an odd integer divided by 2, in 
accordance with Cor. 8. In fact, substituting Newton's 
numerical values  $\beta =11/2$,
one finds that the maximum  rate of advance is $21.57'$, and of retardation 
$14.83'$. This is in reasonable  agreement with the values $23'$ and 
$16$ $1/3'$ given in the original 
(1687) Scholium to Prop. 35
corresponding to $\beta \approx 6$.
In Cor. 9 Newton gave a qualitative
argument for the variability of the eccentricity, but
there is no evidence that he obtained this quantitative result
from his ``theory of
gravity'' .  According to his theory the 
maximum variability of the apogee is $ 15m/8=8^0 2'$ 
instead of $12^0 18'$ as quoted in the Scholium to Prop. 35.
Although the  lunar  model of  Horrocks  was probably the
inspiration for his Portsmouth method, in the end Newton 
was able to account partially for this model from his dynamical
principles.} \fo{\no These anomalies in the orbit of the moon around the earth had been a major challenge to astronomers
 since Antiquity. Already by the second century B.C., Hipparchus had found that the moon's motion
 at quadrature deviated in longitude  by over two and a half degree from the predictions of the Greek
 model of epicyclic motion, although this model accounted for the moon's  position at conjunction and
 opposition from the sun. Subsequently, Ptolemy  proposed the first mechanism to account for this
 anomaly, known as the \un { evection}, but his mechanism also predicted a near doubling  of the apparent
 size of the moon during its orbit which is not observed. Nevertheless, Ptolemy's lunar model was not
 challenged until the 15-th century when the Arab astronomer Ibn-al Shatir develop and alternative
 mechanism for the lunar motion which was later adopted by Copernicus.  Their model accounted
 for the evection without introducing the large unobserved variations of the lunar size in Ptolemy's
 model. In the 17-th century alternative models where developed by Tycho Brahe and Kepler who
 incorporated  his law of areas for planetary motion into his lunar model.  In 1640, Jermy Horrocks
 refined Kepler's model further predicting  correctly the inequalities in the distance of the moon
 from the earth. These are some of the additional  inequalities that Newton was also able to demonstrate
 to be caused by the gravitational force of the sun acting on the moon (Nauenberg 2001b)}.

\enq

 Newton's  lunar work  was received
 with immense admiration by those who were able to understand the profound mathematical
 innovations in his theory. An early reviewer of the second edition of the {\it Principia} stated
 that 
 \beq 
 the computations made of the lunar motions from their own causes, by using the
 theory of gravity, the phenomena being in accord, proves the divine force of intellect and the
 outstanding sagacity of the discoverer
 \enq 
Laplace asserted that the sections of the {\it Principia} dealing with the motion of the moon
are one of the most profound  parts of this admirable work \fo{\no Parmi les in\'{e}galit\'{e}s da mouvement de la Lune en longitude, 
                                   Newton n`a d\'{e}velopp\'{e} que la {\it variation}. La m\`{e}thode  qu' il  suivie me
                                   parait  \^{e}tre une des choses le plus remarquables de l'Ouvrage des {\it  Principes} (Laplace 825, 409)},
and the British Astronomer Royal, George
Airy, declared ``that it was the most valuable chapter that has ever been written on physical
science" (Cohen 1972) .  The French mathematician and astronomer  Fran\c{c}ois F\`{e}lix Tisserand
in his  {\it Trait\'{e} de M\'{e}canique C\'{e}leste} (Tisserand 1894) 
carefully  reviewed  Newton's lunar theory as it appeared
in the {\it Principia},
and also compared some of Newton's
results in the Portsmouth manuscript with the results of the variation of parameters
perturbation theory of Euler, Laplace and Lagrange. For 
an arbitrary perturbing force, Tisserand  found that 
Newton's  equation  for the rotation of the major axis of the ellipse 
 was correct to lowest order in the eccentricity of the 
orbit,  while his application to the lunar case differed 
only in the numerical value of one parameter,
which Newton gave as $11/2$, instead of the correct value of $5$ (Nauenberg 2001a)
In particular, Tisserand  concluded that 
\beq
Newton derives  entirely correctly that the average annual movement of the
apogee is  $38^051'51"$,  while the one given in the astronomical tables  is
$40^{0}41'5"$  \fo{\no Newton d\`{e}duit, tout \`{a} fait correctement ...que le mouvement
moyen annuel de l'apog\`{e}e est de $38^051'51"$, tandis que
celui qui est donne dans les Tables astronomiques es de
$40^{0}41'5"$ (Tisserand 1894, 45)}
\enq

 D'Alembert, however,  doubted whether some of Newton's derivation were really sound, and complained that 
 \beq
 there are some that M. Newton said to have calculated with the theory of gravitation,
 but without letting us know the road that he took to obtain them. Those are the the
 ones like $11' 49"$ that depend on the equation of the sun's center \fo{\no  en est quelques-unes que M. Newton did avoir calcul\'{e}es par la Theorie de la gravitation,
mais nous apprendre le chemin qu'il a pris pour  y parvenir. Telle son celles de 11' 49" qui
d\'{e}pend de l' \'{e}quation du centre du soleil.}.
 \enq
Here, d'Alembert  was referring to  Newton's calculation,  in the Scholium mentioned previously,  of the annual
equation of the mean motion of the moon which depends on the earth's eccentricity $\epsilon$ in
its orbit around the sun. Newton had taken $\epsilon$ equal to 16 7/8 divided by 1000, and 
D'Alembert may have  been aware that the amplitude of this perturbation is  $3\epsilon m= 13' $
where  $m$ is the ratio of the lunar sidereal period to a period of one year.478
 Hence, although  in this Scholium  Newton had stated that his  results had been obtained by ``his theory of gravity",  it appears
  that he adjusted some  of the perturbation amplitudes  to fit the observational data .
 
For the next two centuries after the publication of the {\it Principia},  Newton's approach to what became known as 
the {\it three body problem}\fo {\no 
Given the initial conditions (position and velocities ) for three  bodies moving under the action
of their mutual gravitational attraction, to determine their motion at all times in the future.} in dynamical astronomy 
stimulated the work of  mathematicians and
astronomers,  and this problem  remains a challenge up   to the present time \fo{\no
For example, in 1772 Lagrange discovered an exact solution of the three body problem where each of 
the celestial bodies move in elliptic orbits with the same period,  and  with a common focus located  at their  center of mass.
The stability of these orbits, however, was not  examined fully until much later, first by  Routh  (1875)
for the special case of circular orbits, and later for elliptic orbits  by Danby (1964). These studies  were  restricted
to linear instabilities, and a non-linear instability  analysis  has been undertaken only recently  (Nauenberg 2002).} .
By the late 1700's Lagrange and Laplace had written major
treatises on analytic mechanics (Lagrange 1811), and  celestial mechanics  (Laplace 1878)  containing  the mathematical progress that had
been made. There is an often repeated tale \fo{\no
 Se non \`{e} vero, \`{e} ben trovato.  } that Napoleon once asked Laplace why  God did not appear    
in his work,  and that Laplace famously responded
``I didn't need  that hypotheses", but  in print he declared  that 
\beq
 These phenomena and some others similarly
 explained, lead us to believe that  everything depends on these laws [the primordial laws of nature] by relations more
 or less  hidden, but of which it is wiser to admit ignorance, rather than to substitute imaginary causes solely in order
 to quiet our uneasiness  about the origin of the things that interest us (Morando 1995, 144)\fo{\no
 Ces ph\`{e}nom\'{e}nes et quelques autres semblablement expliqu\'{e}s autorisent \`{a} penser que tous
 d\'{e}pendent de ces lois, par des rapport plus ou moin cach\'{e}s,  mais dont  il es plus sage d'avouer  l'ignorance que
 d'y substituer des cause imagin\'{e}es par le seul besoin de calmer notre inqui\'{e}tude  sur l'origine des choses qui nous
 int\'{e}resent (Laplace 1835, 478).}.
 \enq
Newton  claimed  God needed to interfere from time to time in order to maintain the
stability of the solar system, but Laplace asserted  that he had been able to give a
mathematical proof of this stability. Later, however, this proof was shown to be
flawed \fo {\no Laplace's proof that secular variations of the mean solar  distances of the
planets do not occur were based on perturbation expansions up to third order in the eccentricities,
but these expansion were shown  not to be convergent.}
by the work of Henri Poincar\`{e} (1892).

The overall impact of Newton's {\it Principia} in astronomy was  best summarized by
 Laplace's conclusion,
 \beq
This admirable work contains the germs of all the great discoveries  that have been made since, 
about  { \it the system of the world}: the history of its development by the followers of that great
geometer will be at the same time the most useful comment on his work, as well as
the best guide to arrive at knew discoveries \fo{\no
Cet admirable Ouvrage contient les germes de toutes les gandes  d\`{e}couvertes qui ont
\`{e}t\`{e}s faits depuis  sur le syst\`{e}me de monde: l'histoire de leur d\'{e}veloppement par
les successeurs de ce grand g\'{e}ometr\`{e} serait  \`{a} la fois le plus utile commentaire de son 
Ouvrage, ce le meilleur guide pur arriver  \`{a} de novelles d\`{e}couvertes.} 
\enq

\subsection*{Acknowledgements}
I would like to thank Niccolo  Guicciardini for many valuable comments. 
 For the  Introduction, I am particularly indebted to
M. Feingold's account in \\  {\it The Newtonian Moment,  Isaac Newton
and the making of modern culture} (Feingold, 2004).

\subsection*{Appendix,  Newton's moon test}
                                  Newton's assumption that  the inverse square law for gravitational forces applies on the surface 
                                 on the earth, requires  the relation $a_m/g=(r_e/r_m)^2$ , where $a_m$ is  the radial
                                  acceleration of the moon towards the earth, $g$ is the gravitational
                                 acceleration at the surface of the earth,  $r_m$
                                 is the radius of the moon's orbit, and  $r_e$ is the radius of the earth.  Since Newton had found that
                                  $a_m=4\pi^2 r_m/T_m^2$, he  tested  the inverse square law by calculating the ratio
                                 $(g/ 2)/d_m$,  where $g/2$ is the distance a body falls in one second  on the surface of the earth, and
                                 $d_m=a_m/2=2\pi^2 r_m/T_m $ is the corresponding distance that  
                                 the moon `` descends  towards the earth
                                 in one second".   
                                 Pendulum experiments had established that $g=32$  feet/ $sec^2$, but
                                 to obtain $d_m$, Newton first  had to calculate  
                                 $d_e=2\pi^2 r_e /T^2_e$, which  is the corresponding distance of fall for a body on the surface
                                 of the earth co-rotating with the earth's  period $T_e$ of one day.  
                                 
                                 Taking for the earth's radius
                                 $r_e=3500$ miles, and assuming that a mile is 5000 feet,  he obtained  
                                 $d_e=5/9$ inches, and $(g/2)/d_e= 345.6$  which he rounded to 350.  Huygens  had 
                                 carried out a similar calculation,  but taking  a different value of the earth's radius, $r_e=3711$ miles,
                                  and $g=27.33$ feet/ $sec^2$.  he obtained 
                                 for this ratio the value 265 (Huygens 1929),   while the correct value is  $290$. This result
                                  answered  the long  standing question why,   if
                                 the earth was spining about its axis  once a day, objects on its  surface do not fly off:
                                 \beq
                                 The force 
                                 of gravity is many times greater that what would prevent the rotation of the earth from causing
                                 bodies to recede from it and raise into the air (Herivel 1965, 196)
                                 \enq
                                                                  Since  $d_e/d_m=(r_e/r_m)(T_m/T_e)^2$, where  
                                  $T_m/T_e=27.3216 $, and   $r_m/r_e \approx 60$ which was  already measurd  by   
                                   Greek astronomers,  one obtains  $d_e/d_m$=12.44 ( 
                                 Newton rounded it to 12.5).  Hence, $(g/2)/d_m= 16(9/5)12.5=4320$  which differs  appreciable
                                 from the expected value $(r_m/r_e)^2 \approx 3600$.                                 
                                 Newton's  only comment about this discrepancy was   that the force of gravity at the surface of the Earth                                                                    
                                 \beq
                                
                                 is 4000 and more times greater than he endeavor of the Moon to recede from the Earth,
                                 \enq                                  
                                 but he must have been gravely disappointed with this result.                                  
                                 The reason for the  failure of  Newton's early  {\it moon test}  is that  in his calculations
                                  he had used an incorrect value for the 
                                  radius of  the earth based on a value  of  about 61 English miles per degree of latitude,
                                   and also that he had assumed that a mile  corresponds to 5000 feet instead of the
                                  correct value 5280 ( in this manuscript Newton  stated that `` $1/30$ of a mile is $500/3$ feet"). 
                                  Apparently he
                                  did not become aware of his errors  until 1683,  when he substituted in his relation a much 
                                  better  value for the earth's radius $r_e$ obtained in 1669 by Picard 
                                   from his measurement for  a degree of latitude of 
                                   69.2 English miles (see Prop. 19, Book 3).  This measurement  gives   $r_e=3965$ miles, close to the modern value. 
                                   In this case $(g/2)/d_m=4320 (61/69.2)(5000/5280)= 3606$,
                                  in excellent agreement with the result predicted by Newton's theory.

\begin{list} {}{\itemsep .3in }

\item  Aiton, E. J. (1960) The Celestial Mechanics of Leibniz' {\it Annals of Science} 15: 65-82

\item  Aiton, E. J. (1995) The vortex theory in competition with Newtonian celestial dynamics,
                                    {\it Planetary Astronomy from the Renaissance to the rise of Astrophysics}  edited by 
                                     R. Taton and C. Wilson, Cambridge:  Cambridge University Press.

\item Barrow-Green, June (1991) {\it Poincar\'{e} and the Three Body Problem} American
                                 Mathematical Society, History of Mathematics vol. 11.           

\item Bernoulli, Johann  (1710) Extrait de la R\'{e}sponse de M. Bernoulli \`{a} M. Hermann, dat\'{e}e de
                                     Basle le 7 Octobre 1710,   {\it M\'{e}moires de l'Acad\'{e}mie des Sciences}: 521-533.

\item Bertoloni Meli, Domenico (1991) {\it Equivalence and priority: Newton versus Leibniz} Oxford:Clarendon Press

\item Besterman, T. (1968)  {\it Voltaire Notebooks}, ed. T. Besterman, in {\it The Complete Works of Voltaire} vol. 81 : Toronto: 83

\item Brackenridge, J. Bruce (1995) {\it The Key to Newton's Principia}, Berkeley: University of  California Press.
 
 \item Brackenridge,  J. Bruce  and Nauenberg, Michael (2002) Curvature in Newton's Dynamics,  
                                  in I.B. Cohen and G. E. Smith (eds) {\it The
                                  Cambridge Companion to Newton}.  Cambridge: Cambridge University Press :
                                 85-137.
                             
\item  Cajori, Florian (1934)  {\it Principia} vol. 2 ,  Motte's translation revised 
                                   by F. Cajori, Berkeley: University  of California Press: 664.

\item   Chandrasekhar, S (1995)  {\it Newton's Principia for the Common reader}, Oxford: Clarendon Press

\item   Cohen, I. Bernard (1999)  { \it The Principia, Mathematical Principles of Natural
                      Philosophy}. A new translation by I.Bernard  Cohen
                      and Anne Whitman assisted by Julia Budenz, preceded by a guide to Newton's Principia by I. Bernard Cohen.
                      Los Angeles, Berkeley, London:
                       University of California Press.

\item    Cohen, I. Bernard  (1972)  Historical Introduction to   Isaac Newton's { \it Theory of the Moon's Motion}, 
                           Dawson (1972).

\item  Danby, J. M. A (1964) Stability of Triangular Points in the Elliptic Restricted Problem
                            of Three Particles, {\it The Astronomical Journal} 69: 165-172      

\item    Euler, Leonhard (1769)  { \it Opera Omnia} Series secunda,Opera
                      Mechanica et Astronomica, vol. 23, ed. L. Courvoisier
                    and J.O. Fleckenstein, Basel, MCMLXIX: 286-289.
                    
\item    Euler, Leonhard (1772) `Theoria Motuum Lunae, Nova Methodo Pertractata' { \it Opera Omnia} Series secunda,Opera
                      Mechanica et Astronomica, vol. 22, ed. L. Courvoisier, Lausanne, MCMLVIII.

\item   Euler, Leonhard  (1983) Une  Th\`{e}orie de Saturne et de Jupiter, par laquelle on puisse expliquer le in\`{e}galit\`{e}s
                                 que ces deux Plan\`{e}tes paroissert se causer mutuellement, principalement vers les temps de leur conjoction,
                                 quoted by O. Volk,  Eulers Beitr\"{a}ge zur Theorie der Bewegungen der Himmelsk\"{o}rper , in \\                                                                 
                                 {\it Leonhard Euler, 1707-1783, Beritr\"{a}ge zu Leben und Werke }, Basel:Birkh\"{a}user, pp. 345-361.

\item Feingold, Mordechai (2004) {\it The Newtonian Moment,
                            Isaac Newton and
                       the making of modern culture}, New York, Oxford: Oxford University Press.

\item Fontenelle, Bernard Le Bouvier (1728)  {\it The elogium of Sir Isaac Newton.}  Printed for J. Tonson,   London

\item  Guicciardini, Niccol\`{o} (1999) { \it Reading the Principia, 
                               The debate on Newton's mathematical
                               methods for natural philosophy from 1687 to 1736}, Cambridge : Cambridge University Press.
                              
\item  Herivel, J. (1965) {\it The Background to Newton's Principia, A study of Newton's
 dynamical researches in the years 1664-1684 } Oxford:  Clarendon Press:  192-198.                          

\item   Hermann, Jacob (1710) Extrait d'une lettre de M. Herman \`{a} M. Bernoulli, dat\'{e}e de Pado\"{u}e le 12, Juilliet 1710
                             {\it M\'{e}moires de l'Acad\'{e}mie des Sciences}: 519-521   

\item Hill, George W.  (1878)  ` Researches in the Lunar Theory' \un{ American Journal of Mathematics}
                            1 (1878) 5-26, 129-147, 245-260.  Reproduced in \un{Collected Mathematical Works
                            of G. W. Hill}, vol. 1 Washington:Carnegie Institute (1905) : 284-335.

\item  Hooke, Robert  (1674) An Attempt to Prove the Motion of the Earth from Observations,  reprinted in 
                         Gunther, R. T.   { \it Early Science in
                            Oxford} vol. 3 , Oxford:  Printed for the Author (1931) 1-28
                            
\item Huygens, Christiaan  (1673)  C. Huygens, {\it Horologium Oscillatorum}, (Paris 1673). 
                                   English translation by Richard J. Blackwell,
                                  { \it C. Huygens,  The Pendulum Clock} Ames, 1986

\item  Huygens, Christiaan (1690) {\it  Trait\`{e} de la Lumiere, avec un  Discourse sur  la Cause de la Pesanteur},  Leyden.

\item   Huygens, Christiaan (1897) {\it Ouvres Compl\'{e}tes de Christiaan Huygens} , vol. 7, The Hague,  Martinus Nijhoff,
                                    p. 325
                          
 \item   Huygens, Christiaan (1929) ` De Vi Centrifuga, Appendice  I. 1659 ' \\
             {\it Ouvres Compl\'{e}tes de Christiaan Huygens} , vol. 16,  The Hague,  Martinus Nijhoff, 
                                    p. 304.

\item  Kline, Morris (1972)  {\it Mathematical Thought from Ancient to Modern Times} Oxford : Oxford University Press :554-557
                                    
\item Lagrange, J. L (1811)  { \it M\'{e}canique Analytique} Paris: Coursier, Paris

\item  Laplace, Pierre-Simon (1796)  {\it Exposition du Syst\`{e}me du Monde} Paris: Bachelier

\item    Laplace, Pierre Simon  (1822)  { \it A treatise of Celestial Mechanics}, translated from the French
                               and elucidated with explanatory  notes by the Rev. Henry H. Harte,  Dublin : Printed for R. Milliken 

\item    Laplace, Pierre Simon (1825) {\it Trait\'{e} de  M\'{e}canique C\`{e}leste}  vol. 5,  Paris: Bachelier Libraire

\item    Laplace, PIerre Simon (1835) {\it Trait\'{e} de  M\'{e}canique C\`{e}leste},  vol. 6, Paris: Bachelier Libraire 

\item    Laplace, PIerre Simon (1878) {\it Trait\'{e} de  M\'{e}canique C\`{e}leste},  5 volumes, Paris: Gauthier-Villars. 

\item    Leibniz, Gottfried Wilhelm (1973)   { \it Marginalia in Newtoni Principia Mathematica} edited by E. A. Fellmann, Paris: Vrin

 \item Locke,  John (2001) {\it An essay in human understanding}. Ontario: Batoche Books                     

\item   Lohne, J (1960) Hooke versus Newton, {\it Centaurus} 7: 6-52. 

\item   Morando, Bruno (1995) Laplace, in Rene Taton and Curtis Wilson (eds)\\
 {\it The General History of
Astronomy} 2B,\\  {\it Planetary Astronomy
from the Renaissance to the rise of  Astrophysics,} 
Cambridge: Cambridge  University Press

\item Nauenberg, Michael (1994a)  Newton's Early Computational Method for Dynamics,
                                  {\it Archive for History of Exact Sciences } 46 :221-252

\item  Nauenberg, Michael (1994b)  Hooke, Orbital Motion, and Newton's Principia,

                                         {\it American Journal of Physics}, 62: 331-  350. 

\item Nauenberg, Michael (1994c)  Newton's
                                      Principia and Inverse Square Orbits,
                                      
                                       { \it The College Mathematics Journal} 25: 212-222. 
                                     
\item Nauenberg, Michael (1996) Huygens and Newton on Curvature and its applications
                                  to dynamics  {\it  De zeventiende eeuw, Cultur in de Nederlanden in interdisciplinair 
                                  perspectief}, jaargang 12 : 215-234.

\item Nauenberg, Michael (1998) `The Mathematical Principles underlying the Principia
                                  revisited',
                                   {\it Journal for the History of Astronomy} 29: 301-307.

\item  Nauenberg, Michael (2000) {\it Newton's Portsmouth perturbation Method
                                and its application to lunar motion},  in R.H Dalitz and M. Nauenberg (eds)
                                 {\it The Foundations of Newtonian Scholarship}
                                 Singapore: World Scientific
                                
\item Nauenberg, Michael (2001a) ` Newton's perturbation methods for the three-body
                                 problem and their application to Lunar motion', in  J.Z. Buchwald and I. B. Cohen
                                 (eds)   {\it Isaac's Newton Natural Philosophy},
                                 Cambridge: MIT Press.

\item Nauenberg, Michael   (2001b) `Essay review of N. Kollerstrom's 
                                  Newton's Forgotten Lunar Theory' {\it Journal for the History of Astronomy} 32:
                                  162-168.

\item Nauenberg, Michael  (2002)  ` Stability and eccentricity of periodic orbits for two planets in 1:1 resonance'
                                  {\it The Astronomical Journal} 124:2332-2338
                                                                   
\item  Nauenberg, Michael (2003a) ` Kepler's area law in the Principia: filling in some details in
                                    Newton's proof of Proposition 1 ',  {\it  Historia Mathematica} 30: 441-456.

\item Nauenberg, Michael (2003b) `Orbites p\'{e}riodiques du probl\'{e}me des trois
                                corps: les origines, les contributions de Hill et Poincar\'{e}. et quelques 
                                 d\'{e}veloppements r\'{e}cents' ,  in E. Charpentier, E. GHys and A. Lesne (eds)
                                 {\it L'h\'{e}ritage scientifique de Poincar\'{e}}, Paris:                                 
                                  Belin, pp. 128-157. 

\item  Nauenberg, Michael (2005) `Curvature in Orbital Dynamics'.  {\it  American Journal 
                                             of Physics} 73 : 340-348.

\item Nauenberg, Michael (2005)                                                                              
                                         `Robert Hooke's Seminal Contributions to Orbital Dynamics', 
                                         {\it Physics
                                         in Perspective 7}: 4-34, reproduced in M.  Cooper and M. Hunter (eds)                                         
                                          {\it Robert Hooke,
                                         Tercentennial Studies}, London:
                                          Ashgate ( 2006) pp. 3-32. 

\item Nauenberg, Michael (2010)  ` The early application of the calculus to inverse square problem',
                                           {\it Archive for the History of Exact Science} (currently on line in electronic version,
                                            to appear in May issue )                                         

\item Newton, Isaac (1702) { \it Theory of the Moon's Motion}, with a 
                                  bibliographical and historical introduction by I. B. Cohen,  Dawson (1972) :41.
                                  
\item   Pemberton, H. (1728)  {\it A view of Sir Isaac Newton's Philosophy}  London: 
                                   Palmer.

\item Poincare,   Henri (1892))  {\it  Les M\'{e}thodes Nouvelles de la Mechanique C\'{e}leste},
                                  Paris:Gauthier-Villars.

\item  Routh, E. J. (1875) ` On Lagrange's  Three Particle with a Supplement on the Stability of Steady Motion'
                               {\it Proceedings of the London Mathematical Society} 6: 86-96.  Reprinted in
                               {\it Stability of Motion} edited by A. T Fuller, New York: Halsted Press.

\item  Tisserand,  Felix (1894) `Th\'{e}orie de la Lune de Newton'
                        {\it Trait\'{e} 
                       de M\'{e}canique C\'{e}leste}, tome III,
                       Chapitre III,  Paris:
                       Gauthier-Villars et Fils, pp. 27-45
                                   
\item   Todhunter, I  (1962)  { \it A historty  of the  mathematical  theories of attraction  and the
                                   figure of the earth}, New York : Dover.

\item    Turnbull, H. W (1960)  {\it  The correspondence of Isaac Newton } vol.2, 1676-1678, edited by
                        H.W. Turnbull, Cambridge Univ. Press,   pp. 297-314.
                        
\item   Turnbull , H. W.  (1961)  {\it The Correspondence of Isaac Newton}  vol. 3 ,  1688-1694,  edited by  H. W. Turnbull,
                                   Cambridge: Cambridge University Press,  p. 236. 

\item   Varignon, Pierre  (1701)  `Autre Regles Generale des Force Centrales' { \it M\'{e}moires de l'Academie des Sciences}

\item    Voltaire, Francois-Marie (1734)   {\it Letters on England}; translated by L. Tancock, Harmondsworth:
              Penguin Books, 1984, pp. 71-72 
                                     
\item     Waff, Craig. B. (1976) `Isaac Newton, 
                       the motion of the lunar apogee, and the
                       establishment of the inverse square law',
                         {\it Vistas in Astronomy} 20
                       (1976): 99-103; ``Newton and 
                       the Motion of the Moon: An
                       Essay Review'  {\it  Centaurus} 21 (1977): 64-75.
                       
\item    Waff, Craig. B. (1995)  'Clairaut and the motion of the
lunar apse: the inverse-square law undergoes a test', in Rene Taton and
Curtis Wilson (eds)
{\it The General History of
Astronomy} 2B,   {\it Planetary Astronomy
from the Renaissance to the rise of  astrophysics,} 
Cambridge: Cambridge  University Press, pp. 35-46.

\item   Westfall, R. S (1995)                                                                      
                                  {\it  Never at Rest}, Cambridge : Cambridge University Press,  p. 505

\item  Whiteside, Derek (1967)  { \it The Mathematical Papers of Isaac Newton}, vol. 1,  1664-1666, 
                                 edited by D.T. Whiteside, Cambridge: Cambridge  University Press.

\item  Whiteside, Derek (1974)   {\it The Mathematical Papers of Isaac Newton} 1684-1691,  vol. 6, 
                                 edited by D. T. Whiteside, 
                                 Cambridge:  Cambridge University  Press, pp.  30-91.

\end{list}

  \end{document}